\UseRawInputEncoding
\documentclass[aps,pre,preprint,showpacs,floatfix,unsortedaddress,
  superscriptaddress]{revtex4-2}
\usepackage{epsfig}
\usepackage{latexsym}
\usepackage{amsmath}
\usepackage{amsfonts}
\usepackage{amssymb}
\usepackage{amsbsy}
\usepackage{subfigure}
\usepackage{bm}
\usepackage{color}
\usepackage{dcolumn}
\usepackage{hyperref}

\graphicspath{{./figures/}}

\newcommand{\lla}{\left\langle}
\newcommand{\rra}{\right\rangle}

\newcommand{\black}{\color{black}}

\begin{document}
\title{Excluded volume effects
on tangentially driven active ring polymers}
\author{A. Lamura}
\email[]{antonio.lamura@cnr.it}
\affiliation{
Istituto Applicazioni Calcolo, Consiglio Nazionale delle Ricerche (CNR),
Via Amendola 122/D, 70126 Bari, Italy}
\date{\today}
\begin{abstract}
The conformational and dynamical
properties of active ring polymers are studied by numerical simulations.
The two-dimensionally confined polymer is modeled
as a closed bead-spring chain, driven by tangential forces, put in contact
with a heat bath described by the Brownian multiparticle
collision dynamics. Both phantom polymers and chains comprising
excluded volume interactions are considered for different bending rigidities.
The size and shape are found to be
dependent on persistence length, driving force, and
bead mutual exclusion.
The lack of
excluded volume interactions is
responsible for a shrinkage of active rings when increasing driving force in the
flexible limit
while the presence induces a moderate swelling of chains.
Internal dynamics of flexible phantom active rings shows
activity-enhanced diffusive behavior
at large activity values while, in the case of self-avoiding active chains, it
is characterized by
active ballistic motion not depending on stiffness.
The long-time dynamics of active rings is
marked by rotational motion
whose period scales as the inverse of the applied tangential force, irrespective
of persistence length \textcolor{black}{and beads self-exclusion}.
\end{abstract}
\maketitle
\newpage

\section{Introduction}

Last twenty years registered a growing 
interest towards active matter \cite{marc:13,elge:15,bech:16,wink:17}.
{\black{This is made of out-of-equilibrium interacting units
capable of absorbing energy from their environment and
\textcolor{black}{transforming it into motion}.}}
An interesting example is provided by active 
polymer-like structures where
the presence of active noise and/or internal propulsion,
interacting with deformability,
is responsible for intriguing new phenomena, investigated both theoretically
and numerically 
\cite{elge:09,jiang:14,ghos:14,isele:15,isele:16,eise:17,prathyusha:18,duman:18,anand:18,fogl:19,wink:20,eise:22,phil:22_1,vati:23}.
Nature provides numerous realizations showing how activity is crucial
in determining both structural and dynamical properties.
Among others, actin filaments and microtubules are prototypes of
filamentous structures, subject to local active forces exerted by
biomolecular motors, capable of performing different activities
at the biological level \cite{hiro:98,bach:05}.
For example, microtubules placed on kinesin motility assays
can undergo active self-organization to obtain
more ordered structures such as bundles \cite{kawa:10} and rings
\cite{kawa:08}.
Such closed structures are very common and
can be observed in chromosomes
inside bacteria \cite{wu:19}, in DNA and RNA arranging in loops
\cite{schl:92,alex:20}, in actomyosin rings \cite{pear:18}, and
in microtubules on dynein-coated
surfaces \cite{ito:14}
whose dynamics is greatly affected by the circular form \cite{keya:20}.

Very recently some studies have investigated
structures and dynamic behaviors of active rings. 
In three spatial dimensions active Brownian 
\cite{mous:19} and tangentially driven 
\cite{loca:21,phil:22} ring polymer models have been considered.
In the former case it is found that the action of local active random forces
enhances conformal fluctuations \cite{mous:19} while in the latter one, the
local tangent force causes small rings to swell and large rings to collapse
with an arrested dynamics in the case of flexible rings \cite{loca:21}.
Neglecting excluded volume interactions allows an analytical study
of the dynamics of semiflexible
active polar ring polymers \cite{phil:22} which reveals that
conformations are independent on activity and characterized by
a rotational motion. This resembles the tank-treading motion observed
for passive rings \cite{chen:13,lang:14,lieb:18,lieb:20}
and vesicles \cite{nogu:04,beau:04} when subject to an external shear flow.
The interplay of local polar and long-range activities on the swelling and
collapse  of flexible ring polymers has been also considered \cite{kuma:23}.

In the two-dimensional case very few studies addressed the behavior
of active ring polymers. Active Brownian models have been adopted 
to mimic mammalian cells \cite{teix:21} and to investigate the motion
of active rings in porous media \cite{thee:22}. Despite of this, the
problem is interesting since several experiments showed that it is
possible to assemble microtubules, on a motor protein-fixed
surface \cite{kawa:08,liu:11,keya:20}, in ring shapes which are
characterized by rotational motion \cite{ito:14}. Due to the peculiar
dynamic behavior, it appears very engaging to understand such patterns
which strongly depend on the topological constraints in two dimensions. 
This is precisely the aim of the present study where the effects of excluded
volume interactions are explicitly considered in the case of active 
polymer rings.

We introduce a discrete model of a closed semiflexible polymer 
whose beads are subject to a force
tangentially oriented with respect to the polymer backbone.
Excluded volume interactions are taken into account in order
to highlight their roles in the observed dynamics
since these forces are known to be relevant
in the case of two-dimensional
passive rings in the limit of small bending rigidity \cite{saka:10,drub:10}.
Hydrodynamic interactions are
ignored due to the strong interaction
between rings and substrates in two dimensions thus allowing
the use of the free-draining approximation.
For this reason the polymer is placed in contact
with a Brownian heat bath and its dynamics is numerically studied
by using the Brownian version
\cite{ripo:07} of the multiparticle collision dynamics \cite{kapr:08,gomp:09}.
We find that the size and shape, measured by the radius of gyration and by
the asphericity, respectively,
depend on persistence length, excluded volume interactions,
and active force. In the limit of flexible rings,
phantom chains decrease in size when
increasing activity while rings with conserved topology present a moderate
swelling, becoming more roundish in both cases. In the opposite limit of
stiff rings, excluded volume interactions are not crucial
in determining conformations which are independent on activity. 
Flexible phantom active rings show enhanced diffusive dynamics
while self-avoiding active
chains display ballistic dynamic behavior not depending
on stiffness.
The long-time dynamics is characterized by a
reptation motion for all bending rigidities
which, in the case of stiff rings, resembles the tank-treading motion
observed for two-dimensional sheared vesicles \cite{fink:08,kaou:11,lamu:22}.
The rotational period is found to scale as the inverse of the active force.

The numerical model for the polymer and the Brownian heat bath is introduced
in  Sec.~\ref{sec:model}. The results for the 
conformations and the dynamics are reported in Sec.~\ref{sec:results}.
Finally, Sec.~\ref{sec:conclusions} is devoted to discuss the main findings
presenting some conclusions.

\section{Model and Method} \label{sec:model}

A closed chain of length $L$
is considered in two spatial dimensions. It is
composed of $N$ beads, each having mass $M$, whose
internal interactions are due to different contributions.
Consecutive beads interact via the harmonic potential 
\begin{equation}\label{bond}
U_{bond}=\frac{\kappa_h}{2} \sum_{i=1}^{N}
(|{\bm r}_{i+1}-{\bm r}_{i}|-l)^2 ,
\end{equation}
where {\black{$\kappa_h$ is the spring constant,}}
${\bm r}_i$ indicates the position vector of the $i-$th bead
($i=1,\ldots,N$) with ${\bm r}_{N+1}={\bm r}_1$
and ${\bm r}_{0}={\bm r}_N$, and $l$ is the average bond length.
A bending potential is considered to enforce
chain stiffness and is given by
\begin{equation}
U_{bend}=\kappa \sum_{i=1}^{N} (1-\cos \theta_{i})
\label{bend}
\end{equation}
where $\kappa$ controls
the bending rigidity and $\theta_{i}$ is the angle between
two consecutive bond vectors.
In the following, chain stiffness is characterized
in terms of the length $L_p=2 \kappa l/ k_B T$
which corresponds to the polymer persistence
length in the worm-like chain limit \cite{wink:94}.
Here $k_B T$ is the thermal energy, $T$ is the temperature, and $k_B$ is
Boltzmann's constant.
Excluded volume interactions between non-bonded beads are
modeled by the truncated and shifted Lennard-Jones potential
\begin{equation}
U_{ex} =
4 \epsilon \Big [ \Big(\frac{\sigma}{r}\Big)^{12}
-\Big(\frac{\sigma}{r}\Big)^{6} +\frac{1}{4}\Big] \Theta(2^{1/6}\sigma -r) ,
\label{rep_pot}
\end{equation}
{\black{where $\epsilon$ is the volume-exclusion energy,}} $r$ is the distance between two non-connected beads, and
$\Theta(x)$ is the Heaviside function ($\Theta(x)=0$ for $x<0$ and
$\Theta(x)=1$ for $x \ge 0$). This potential avoids
chain self-crossings so to preserve the ring topology.

Finally, an active force ${\bm F}_i^a$ ($i=1,\ldots,N$) is applied tangentially
to the filament at the position of each bead.
In the present paper we adopt a push-pull type force 
\cite{jiang:14,isele:15,anand:18,phil:22}.
\textcolor{black}{By assuming that molecular motors are
homogeneously distributed along a bond,
it is reasonable to consider that each bond is subject to a constant force,
along its direction, given by
$f^a ({\bf r}_{i}-{\bf r}_{i-1})/l (i=1,\ldots,N)$ \cite{jiang:14}.
This force has magnitude $f^a$ since
the bond length $|{\bf r}_{i}-{\bf r}_{i-1}|$
is constrained to be $l$ by using a very high value of the spring constant
$\kappa_h$ in (\ref{bond}). The force on each bond is then equally
distributed between the adjacent beads so that, say, on the bead $i$ there
is a contribution $f^a ({\bf r}_{i}-{\bf r}_{i-1})/(2l)$ along the inward bond
and a contribution $f^a ({\bf r}_{i+1}-{\bf r}_{i})/(2l)$ along the outward
bond. The total net force acting on the $i$-th bead is the sum of these two
terms
\begin{equation}
{\bm F}_i^a =
\frac{f^a}{2 l} \left ( {\bm r}_{i+1}-{\bm r}_{i-1} \right ) , i=1,\dots,N.
\label{force}
\end{equation}
}
The expression (\ref{force}) is such
that the sum of active forces along the discrete ring,
$\sum_{i=1}^{N} {\bm F}_i^a$, is zero \cite{phil:22}.
Moreover, the value of the force (\ref{force}) 
depends on the relative positions of the beads $i-1$ and $i+1$,
varying between $0$, when the two consecutive bonds are antiparallel,
and $f^a$, when the
bonds are parallel.
In other studies a constant tangent force, acting on all the beads,
has been considered \cite{bianco:18,loca:21,mira:23}.
The strength of the total active force is quantified by 
the P\'eclet number
$Pe = f^a N L / (k_B T)$ \textcolor{black}{\cite{isele:15,phil:22}}.
\textcolor{black}{An alternative definition
of the P\'eclet number, $Pe^*= f^a l / (k_B T) = Pe / N^2$,
being $L=Nl$,
is sometimes used in the literature \cite{loca:21}.}
Newton's equations of motion of beads are integrated by
the velocity-Verlet algorithm with time step $\Delta t_p$
\cite{swop:82,alle:87}.

The ring is kept in contact with a Brownian heat bath which is
modeled by making use of the Brownian multiparticle collision (MPC) method
\cite{kiku:03,ripo:07,gomp:09} where hydrodynamics is ignored. 
Every bead \textcolor{black}{interacts}
with \textcolor{black}{$\rho$ virtual solvent particles of mass
$m$}
in order to simulate the interaction with a
fluid volume.
\textcolor{black}{Since it is not necessary to keep track of
the positions of the solvent particles in the present algorithm \cite{ripo:07},
it is sufficient to couple each
bead with an effective virtual solvent particle with momentum
sampled from a Maxwell-Boltzmann distribution of variance
$\rho m k_B T$ and zero mean.}
The interaction process
proceeds via the stochastic rotation dynamics
of the MPC method \cite{ihle:01,lamu:01,gomp:09}.
The relative velocity of each polymer bead, with respect to the
center-of-mass velocity of the bead and its corresponding virtual
solvent particle, is randomly rotated by angles $\pm \alpha$. 
Collisions are then
executed at time intervals $\Delta t$, with   $\Delta t > \Delta t_p$. 
\textcolor{black}{It has been shown that
the evolution equation 
of the MPC model for the solute particle takes the form of a discretized
Langevin equation for which the expression of the friction 
coefficient has been obtained \cite{kiku:03}.}

Simulations are carried out with the choices $\alpha=130^{o}$,
$\Delta t=0.1 t_u$, with time unit $t_u=\sqrt{m l^2/(k_B T)}$,
\textcolor{black}{$M= \rho m$ with
$\rho=5$}, $\kappa_h l^2/(k_B T)=10^4$,
$\sigma/l=1$, $N=L/l=50$,
and $\Delta t_p=10^{-2} \Delta t$.
\textcolor{black}{In some cases, longer rings with $N=100$,
\textcolor{black}{$200$ beads} have been
also considered.}
\textcolor{black}{A larger value of the ratio $\sigma/l$, 
which might be experimentally relevant, would cause the overlap
of neighboring beads with a smoothing of the interaction potential and,
eventually, only minor quantitative changes in the following results.}
The value of
$\kappa_h$ is such to ensure that bond length fluctuations are
negligible in any non-equilibrium condition.
 
\section{Numerical results} \label{sec:results}

We consider rings with persistence lengths
ranging from the flexible limit ($L_p/L=0$) to the stiff one ($L_p/L=40$).
The active force $f^a$ is varied
to access a wide interval of P\'eclet number ($0 \leq Pe \leq 5 \times 10^4$
\textcolor{black}{- $0 \leq Pe^* \leq 20$}).
Finally, 
in order to incorporate excluded volume effects,
the value $\epsilon=k_B T$ is used referring to the model
as a self-avoiding active ring (SAR). To point up topological effects,
a comparison
with self-crossing rings is also carried out
by setting $\epsilon=0$. In this latter case we refer to the model
as a phantom active ring (PAR).
\textcolor{black}{For the considered set of parameters,
the friction coefficient $\xi$ \cite{kiku:03} 
acting on each bead is
such that $M/\xi \lesssim 2.0 \times 10^{-6} \tau_r, 8.5 \times 10^{-5} \tau_r$
for self-avoiding and phantom rings, respectively.
This ensures that the dynamics is close to the overdamped one so that inertial
effects are negligible for the results in the following.
Here and in the rest of the paper, $\tau_r$ denotes the polymer relaxation time in the passive case
and is determined by considering the time decay of the ring-diameter
autocorrelation function (see details when discussing Fig.~\ref{fig:re_bis}).
It results to be $\tau_r \simeq 6.5 \times 10^4 t_u, 1.5 \times 10^3 t_u$
for self-avoiding and phantom flexible rings, respectively, and
$\tau_r \simeq 1.6 \times 10^5 t_u$ when $L_p/L=40$ where there are no
differences between the two models.}
Polymers are initialized in a circular shape
and equilibrated
up to time $10^6 t_u$, much longer than any polymer relaxation time.
Then, data are collected \textcolor{black}{in single runs for every parameter set
over time intervals of duration $\simeq 50 \tau_r$,} and averaged.
\textcolor{black}{In the case of the PAR
model with $L_p/L=0.4$ at $Pe = 2.5 \times 10^4$,
averages are obtained 
from three different realizations, each of duration
up to $150 \tau_r$.}

\subsection{Polymer conformations} \label{sec:conf}

By varying activity and stiffness, rings can attain
different configurations.
In order to characterize the observed patterns, the gyration tensor
\begin{equation}
G_{\alpha\beta}=\frac{1}{N}\sum_{i=1}^{N}  \Delta r_{i,\alpha}
\Delta r_{i,\beta} 
\label{eq:tens}
\end{equation}
is computed. Here
$\Delta r_{i,\alpha}$ is the position of the $i$-th bead
in the center-of-mass reference frame of the polymer and the Greek index
indicates the Cartesian component.
The two eigenvalues $\lambda_1$
and $\lambda_2$, with $\lambda_1 > \lambda_2$, of the tensor (\ref{eq:tens})
are extracted
to calculate the gyration radius
\begin{equation}
R_g^2=\lambda_1+\lambda_2
\label{eq:gyr}
\end{equation}
which measures the total size of the ring. The asphericity
\begin{equation}
A=\frac{(\lambda_1 - \lambda_2)^2}{(\lambda_1 + \lambda_2)^2} 
\label{eq:asph}
\end{equation}
is also computed to provide information about the shape, being
$0 \leq A \leq 1$ with $A=0$ for a circle and $A=1$ for a rod.

\begin{figure}[ht]
\begin{center}
\includegraphics*[width=.9\textwidth,angle=0]{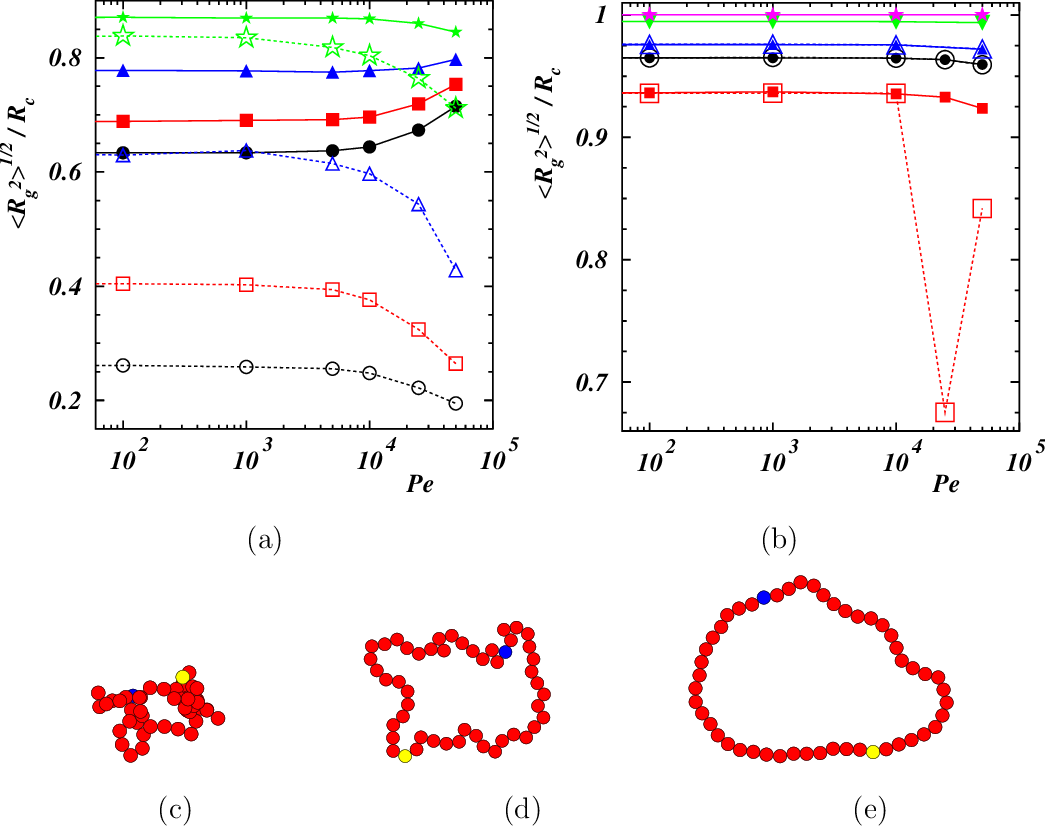}
\caption{Root-mean-square values of the radius of gyration $R_g$
as function of the P\'eclet number $Pe$ in the (a)
flexible regime for $L_p/L= 0$
(black circles), $0.04$ (red squares), $0.1$ (blue up triangles),
$0.2$ (green stars), and in the (b) stiff regime for $L_p/L= 0.4$
(red squares), $0.7$ (black circles), $1$ (blue up triangles),
$4$ (green down triangles), $40$ (purple stars). Filled and empty
symbols correspond to the SAR and PAR models, respectively.
$R_c=L/(2 \pi)$ is the radius of
the corresponding rigid ring.
\textcolor{black}{Typical conformations of rings at $Pe=2.5 \times 10^4$
for the PAR model with $L_p/L=0.04$ (c) and the SAR model
with $L_p/L=0.04$ (d), $0.4$ (e).
The beads $1$ and $N/2+1$ are colored blue and yellow, respectively.}
\label{fig:radgir2}
}
\end{center}
\end{figure}

\begin{figure}[ht]
\begin{center}
\includegraphics*[width=.9\textwidth,angle=0]{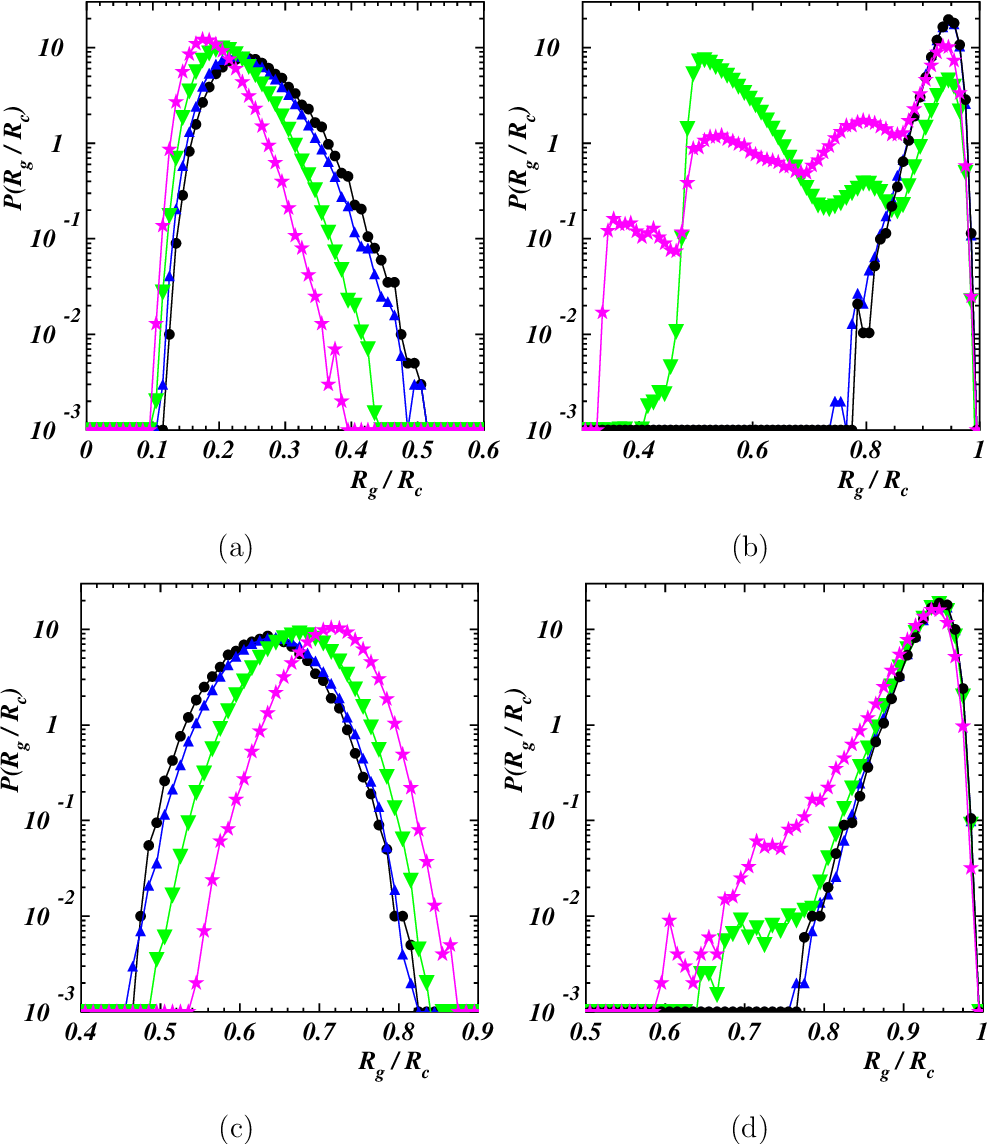}
\caption{Normalized probability distribution function of the radius of gyration
$R_g$ for PAR (upper row) and SAR (lower row) models
with $L_p/L=0$ (a,c), $0.4$ (b,d) 
and $Pe=0$ (black circles), $10^4$ (blue up triangles), $2.5 \times 10^4$
(green down triangles), $5 \times 10^4$ (purple stars).
$R_c=L/(2 \pi)$ is the radius of
the corresponding rigid ring.
\label{fig:pdfrg_pe}
}
\end{center}
\end{figure}

The computed values of $\langle R_g^2 \rangle^{1/2}$, normalized to the
radius of gyration $R_c=L/(2 \pi)$ of a rigid circle,
are depicted versus the P\'eclet number in Fig.~\ref{fig:radgir2}
for different values of the persistence length $L_p$ in the case of SAR
and PAR models. 
The left panel shows data in the flexible regime
corresponding to chains
for which the values of the gyration radius in the
passive limit, $Pe \rightarrow 0$,
are different for self-avoiding (filled symbols) and phantom
(empty symbols) rings \cite{drub:10}.
The difference in the radii 
is due to the conserved circular topology in the SAR model
thanks to self-avoiding effects. In this model
polymers show larger sizes with respect to the PAR model.
On the contrary, the bonds of phantom rings overlap to maximize
\textcolor{black}{the configurational}
entropy because of flexibility \cite{drub:10}
thus producing more compact structures.
Radii increase with the persistence length in both models while the
relative differences reduce.
Activity does not produce any significant change in the radius of gyration up 
to $Pe \simeq 10^3$. For values $Pe \gtrsim 10^4$, 
the behavior varies with the considered model and the conformations depend
on activity. \textcolor{black}{Some typical configurations are reported
in the bottom part of Fig.~\ref{fig:radgir2}.}
This latter range of activity is experimentally
relevant: For example, in the case of
microtubules of length $L=1 \mu m$ with $N=10$ active motors, each 
with force $f^a= 6 pN$,  it would be $Pe \simeq 10^4$ at room temperature
\cite{rupp:12,phil:22}.
Phantom rings tend to shrink while self-avoiding rings
swell. In the case of fully flexible chains ($L_p/L=0$) when
$Pe=5 \times 10^4$, the root mean-square
radius of gyration reduces
by approximately $25\%$ for PAR model and
increases by approximately $15\%$ for SAR model
with respect to the values at equilibrium.
We note here that the shrinkage of phantom chains in two dimensions
is larger compared to the value ($\simeq 10\%$) found in three dimensions
\cite{phil:22} using a similar discrete model for the same P\'eclet number,
thus pointing out
the relevance of space dimensionality on conformations. 
The probability distribution functions $P(R_g/R_c)$
of the radius of gyration are shown for PAR and SAR models with $L_p/L=0$
in the panels (a) and (c), respectively, of Fig.~\ref{fig:pdfrg_pe}
for different values of activity. In both
\textcolor{black}{models,} the mode
of the distribution increases with $Pe$ and the width
becomes narrower suggesting that fluctuations are suppressed by activity
(see Movie 1 given in the supplementary material).
By increasing the stiffness, 
the variations of $\langle R_g^2 \rangle^{1/2}$
with respect to the equilibrium value reduce
and become negligible in the case of self-avoiding
rings for which a very small contraction ($\simeq 3\%$) can be appreciated
when $L_p/L=0.2$. 
At value of bending rigidity such that $L_p/L \simeq 0.4$,
the stiff regime is entered.
\textcolor{black}{In the passive limit $Pe \rightarrow 0$,}
the values of the gyration radius appear indistinguishable
at fixed bending rigidity,
irrespective of excluded volume
interactions, \textcolor{black}{as a consequence of the}
mechanical constraints exerted by stiffness
(see Fig.~\ref{fig:radgir2} (b)).
The global size of rings increases with stiffness to become
comparable to that of a rigid ring for very stiff chains ($L_p/L=40$).
\textcolor{black}{When active}
polymers \textcolor{black}{are considered, they}
show negligible variations in size 
except in the case
of phantom active rings with $L_p/L=0.4$.
\textcolor{black}{In this latter case,}
the gyration radius displays a non-monotonic dependence
on the P\'eclet number due to different
conformations which can be assumed by the ring.
This is reflected in the probability distribution function of $R_g$, shown
in Fig.~\ref{fig:pdfrg_pe} (b), that becomes multimodal in the cases
with $Pe=2.5 \times 10^4, 5 \times 10^4$.
Without the topology constraint
enforced by excluded volume interactions, activity is able to 
deform the chain despite its bending rigidity. The interplay with fluctuations
produces different configurations of variable duration, observable
during very long time dynamics.
Typical patterns, corresponding to the three peaks
of $P(R_g/R_c)$ with $Pe=2.5 \times 10^4$,
are illustrated in Fig.~\ref{fig:conf}.
In the case of self-avoiding active rings
with $L_p/L=0.4$, activity does not change the global size. However,
distribution functions become skewed (see Fig.~\ref{fig:pdfrg_pe} (d))
since rings continuously shrink and swell
during their dynamics (see Movie 2 given in the supplementary material).
This effect reduces when increasing the bending rigidity so that rings
behave as rigid circles.
Indeed, when $L_p/L \gtrsim 1$,
no appreciable difference can be observed in the behavior between
PAR and SAR models since self-exclusion becomes irrelevant. This is due to
the fact that bonds are separated from each other because of the high
bending rigidity of stiff polymers. 
More details about the dynamics will be provided in the following Section.

\begin{figure}[ht]
\begin{center}
\includegraphics*[width=.4\textwidth,angle=0]{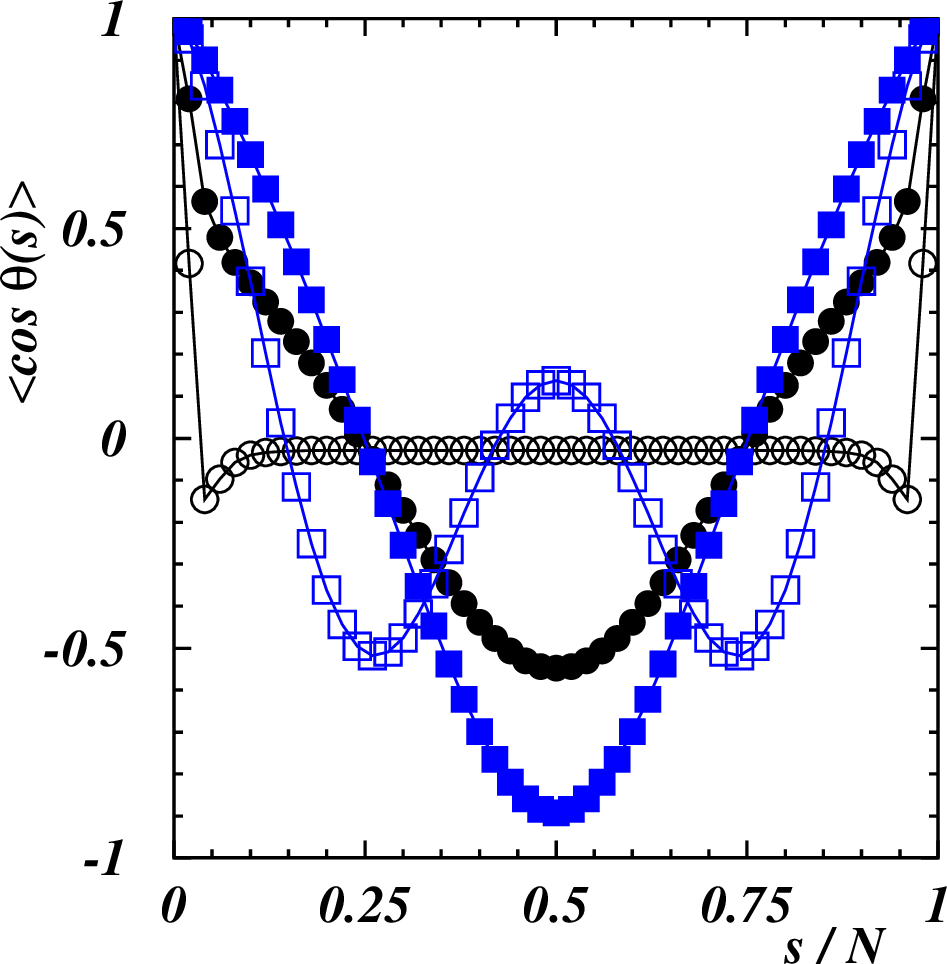}
\caption{Bond correlation function versus the scaled contour separation
$s/N$
for $Pe=2.5 \times 10^4$ and $L_p/L=0$ (black circles), $0.4$ (blue squares).
Filled and empty
symbols correspond to the SAR and PAR models, respectively.
\label{fig:tancorr}
}
\end{center}
\end{figure}

In order to gain further insight into the observed patterns of active rings,
the equal time bond correlation function is computed. It is defined as
\begin{equation}
\langle \cos \theta(s) \rangle =
\frac{\langle {\bm t}_{i+s} \cdot {\bm t}_i \rangle} {l^2}
\end{equation}
where ${\bm t}_i={\bm r}_{i+1}-{\bm r}_i$ is the bond vector
and $s$ is the contour separation. The closed topology guarantees
the property
$\langle \cos \theta(s) \rangle=\langle \cos \theta(N-s) \rangle$.
Figure \ref{fig:tancorr} depicts the bond correlation function for the
persistence lengths $L_p/L=0, 0.4$ with $Pe=2.5 \times 10^4$.
Flexible phantom rings show a very fast decay at small separations
followed
by anti-correlation \textcolor{black}{on a distance of about two bonds}
before reaching complete decorrelation at a contour
separation of about $6$ bonds.
 \textcolor{black}{
This suggests the presence
of small wraps of few beads 
that favor the contraction in size.}
In contrast, flexible
self-avoiding active rings manifest a larger directional correlation
on short distance due to excluded volume effects that restrict the possible
conformations. Owing to the preserved circular topology,
the correlation function
becomes negative on separations $s/N \simeq 1/2$. As already observed, stiffness
is responsible of increasing the size of rings. In the case of self-avoiding
active rings with $L_p/L=0.4$,
this produces a larger correlation between bonds which 
are strongly anti-correlated on distances $s/N \simeq 1/2$ as in the case
of rigid passive rings \cite{saka:10}. When considering
semiflexible phantom active rings,
the presence of the structure with two interlaced rings, shown
in Fig.~\ref{fig:conf} (c),
determines bond anti-correlation at separations
$s/N \simeq 1/4$ and small correlation at $s/N \simeq 1/2$.

\begin{figure}[ht]
\begin{center}
\includegraphics*[width=.9\textwidth,angle=0]{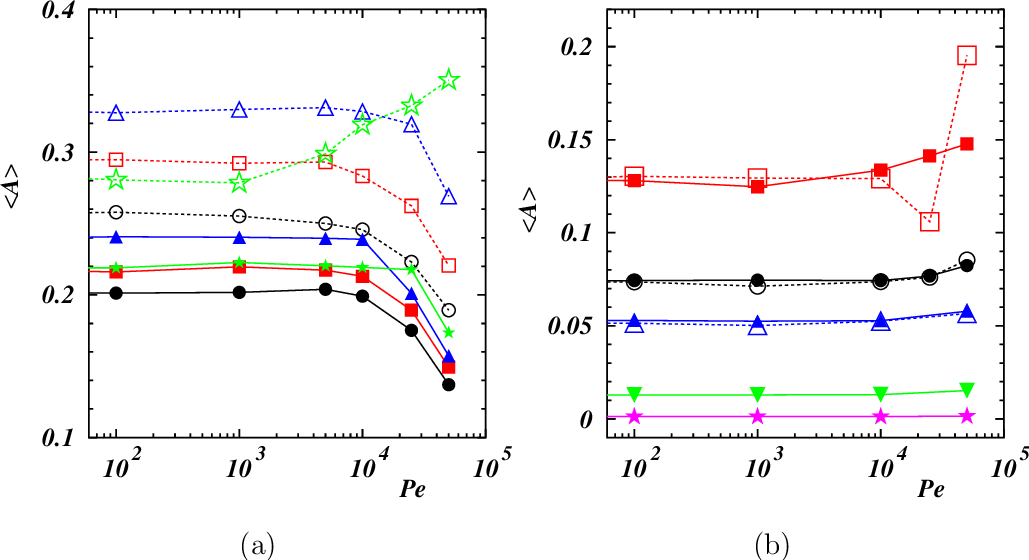}
\caption{Average values of the asphericity $A$
as function of the P\'eclet number $Pe$ in the (a)
flexible regime for $L_p/L= 0$
(black circles), $0.04$ (red squares), $0.1$ (blue up triangles),
$0.2$ (green stars), and in the (b) stiff regime for $L_p/L= 0.4$
(red squares), $0.7$ (black circles), $1$ (blue up triangles),
$4$ (green down triangles), $40$ (purple stars). Filled and empty
symbols correspond to the SAR and PAR models, respectively.
\label{fig:aspher}
}
\end{center}
\end{figure}

\begin{figure}[ht]
\begin{center}
\includegraphics*[width=.9\textwidth,angle=0]{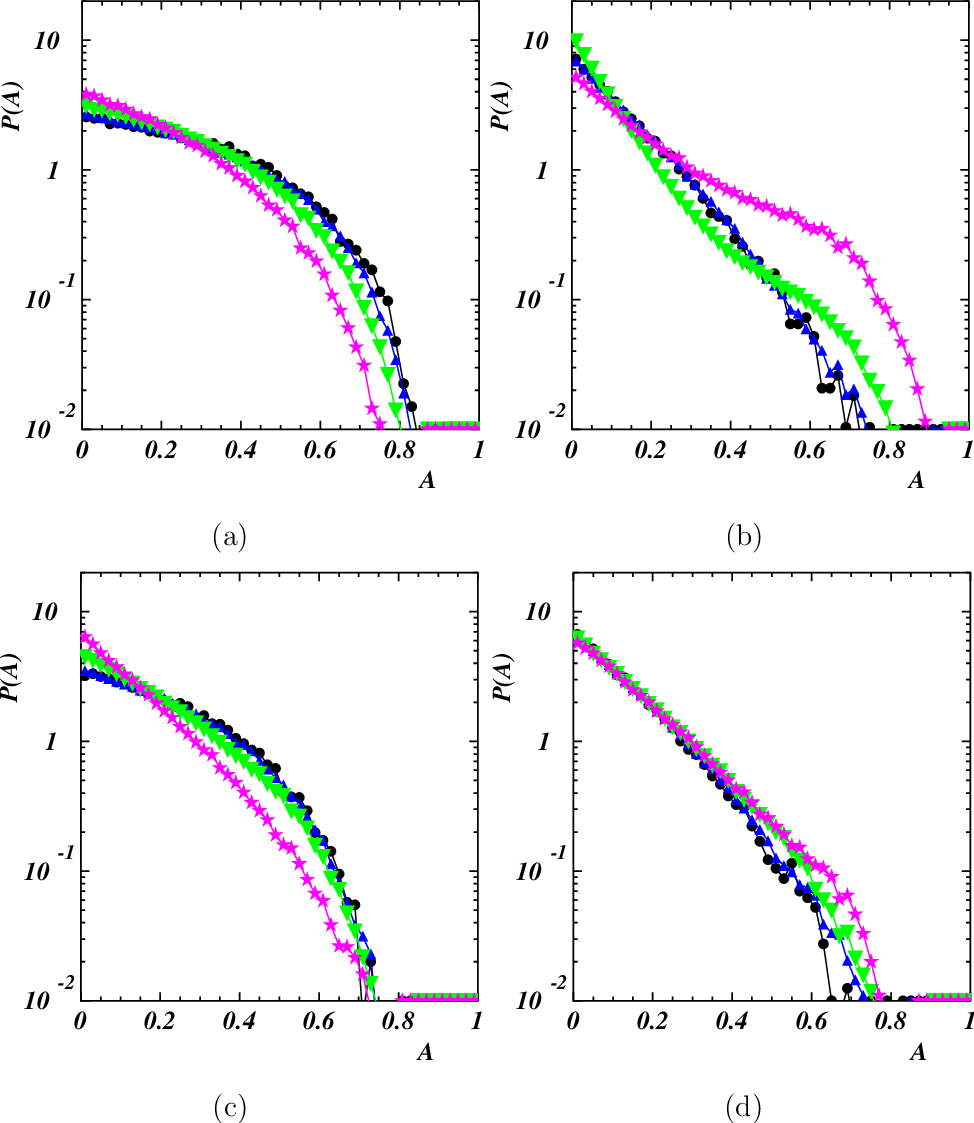}
\caption{Normalized probability distribution function of the asphericity
$A$ for for PAR (upper row) and SAR (lower row) models
with $L_p/L=0$ (a,c), $0.4$ (b,d) 
and $Pe=0$ (black circles), $10^4$ (blue up triangles), $2.5 \times 10^4$
(green down triangles), $5 \times 10^4$ (purple stars).
\label{fig:pdfasf_pe}
}
\end{center}
\end{figure}

In order to better evaluate the effect of activity on the shape of active rings,
the average asphericity is plotted in Fig.~\ref{fig:aspher}
for the flexible (panel (a)) and stiff (panel (b)) regimes.
In the \textcolor{black}{former case,}
asphericity
presents a non-monotonic
dependence on stiffness when $Pe \rightarrow 0$,
\textcolor{black}{as observed in Ref.~}\cite{drub:10}, with
self-exclusion warranting more circular shapes. The effect of activity
is to make rings more roundish in both models
with the exception of the PAR model with $L_p/L=0.2$ when activity favors
elongated structures \textcolor{black}{with respect to the passive limit.
As far as the bending rigidity is
negligible,
\textcolor{black}{our results give $\lla A \rra \simeq 0.26$ in the passive case,
as predicted in the Gaussian limit \cite{dieh:89}.
The observed small wraps at high activity favor
local back folding so that rings are able to gain
even more compact conformations (see Fig.~\ref{fig:radgir2} (a)),
while reducing their asphericity with respect
to the passive case.}
Once bending rigidity comes into play
(at values $L_p/L \simeq 0.2$),
phantom active rings can still reduce the gyration radius due to self-crossing
while assuming a more eccentric elliptical shape.}
The corresponding
probability distributions $P(A)$ are highly skewed with a maximum
at $A=0$ and long tails, as it can be seen in
Fig.~\ref{fig:pdfasf_pe} (a,c) for flexible rings ($L_p/L=0$). 
The effect of activity is to increase the height of the maximum
of distributions while slightly shortening tails.
For stiff active rings (Fig.~\ref{fig:aspher} (b)) it is possible
to observe that activity induces
\textcolor{black}{slightly} more elongated shapes
with respect to the passive
case though this effect reduces when increasing stiffness. Only for
phantom active rings with $L_p/L=0.4$, a non-monotonic dependence on activity
is visible due to the observed conformations (see Fig.~\ref{fig:conf}) and the
peculiar dynamics,
as previously discussed. This is also reflected in the
probability distributions shown Fig.~\ref{fig:pdfasf_pe} (b) for
$L_p/L=0.4$.
The distribution $P(A)$ is characterized by a linear decay as far as
$Pe \lesssim 10^4$. For larger values of activity longer tails
and pronounced shoulders appear in the distribution $P(A)$.
In the case of self-avoiding active rings
(Fig.~\ref{fig:pdfasf_pe} (d)), the role played by activity is to produce
slightly longer tails while poorly affecting the behavior
at small values of $A$.

\subsection{Dynamical behavior} \label{sec:dynamics}

\begin{figure}[ht]
\begin{center}
\includegraphics*[width=.8\textwidth,angle=0]{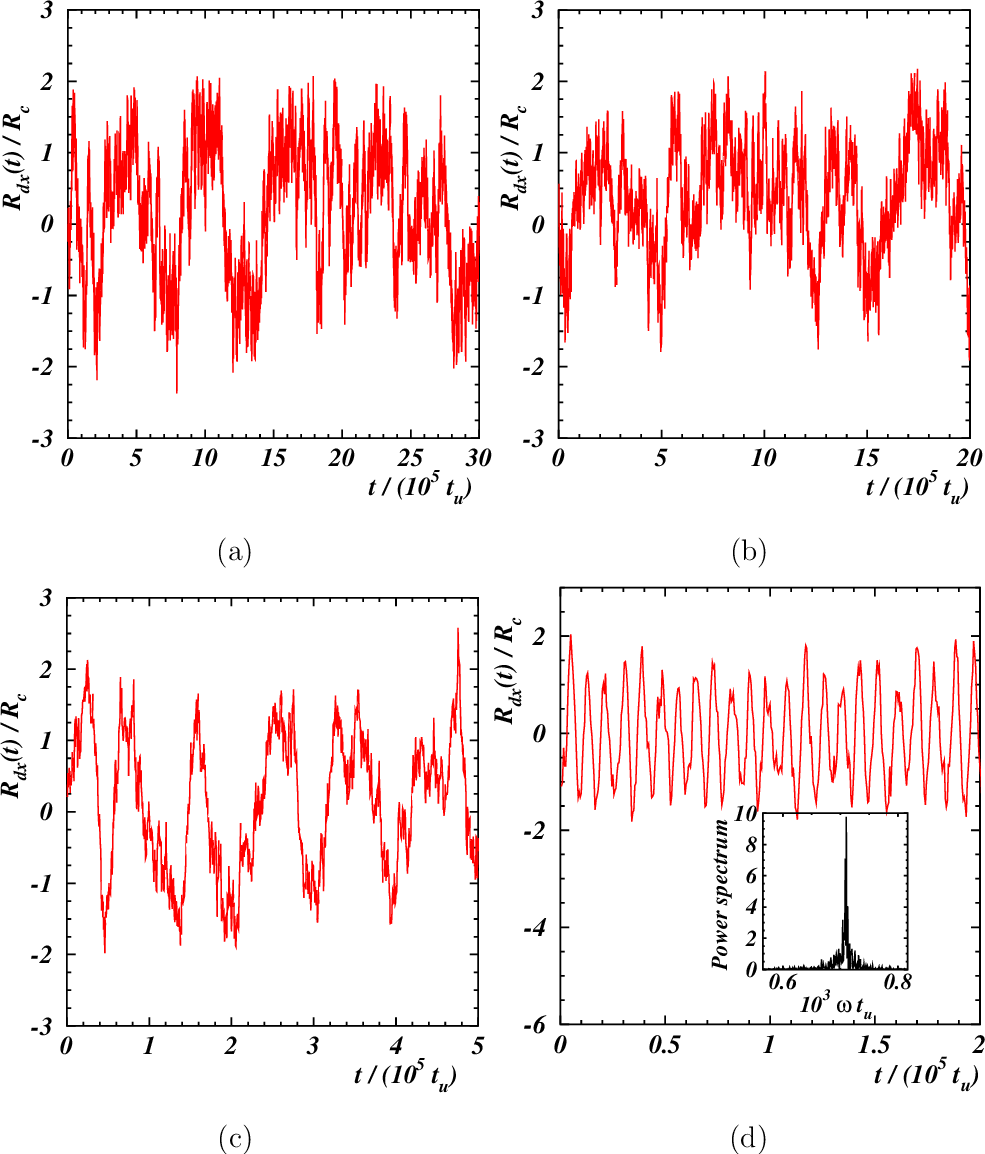}
\caption{\textcolor{black}{Time behavior of the $x$-component $R_{dx}$
of the ring-diameter vector  $\mathbf{R}_d$
for the flexible, $L_p/L = 0.04$, self-avoiding ring
at
$Pe = 0$ (a), $10$ (b), $10^2$ (c), $10^3$ (d).
In the inset of panel (d)
the corresponding power spectrum (in arbitrary units) is shown as a
function of the frequency $\omega$.
$R_c=L/(2 \pi)$ is the radius of the corresponding rigid ring.}
\label{fig:rdx}
}
\end{center}
\end{figure}

\begin{figure}[ht]
\begin{center}
\includegraphics*[width=.4\textwidth,angle=0]{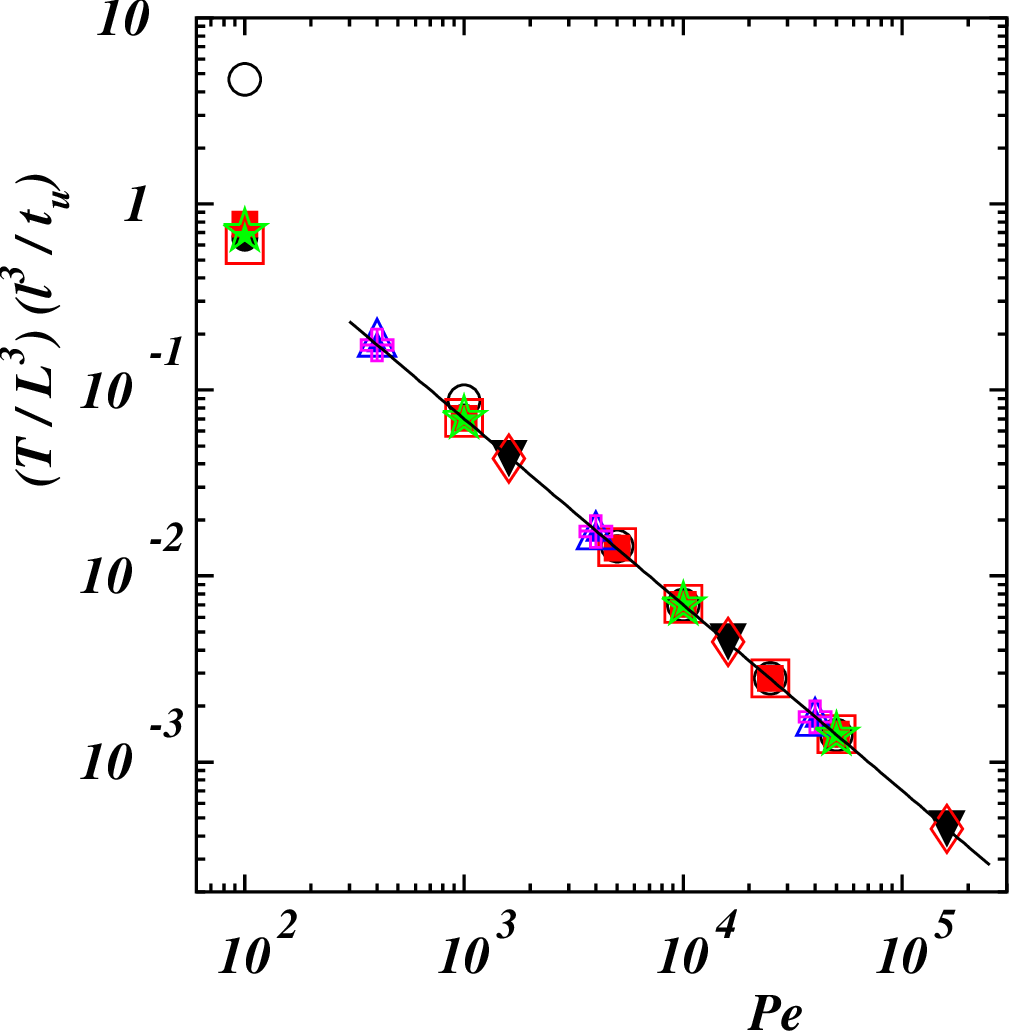}
\caption{Period $T$ of the rotational motion of active rings,
\textcolor{black}{rescaled by the
polymer length $L$,
as function of the P\'eclet
number $Pe$ for $L_p/L=0$ (black circles), $0.1$ (red squares), $1$
(green stars) with $L=50 l$, 
for $L_p/L=0.1$ (blue triangles up), $1$
(purple plus symbols) with $L=100 l$, and
\textcolor{black}{for $L_p/L=0.1$ (red diamonds), $1$
(black triangles down) with $L=200 l$}}. Filled and empty
symbols correspond to the SAR and PAR models, respectively.
The black line has slope $-1$. 
\label{fig:period}
}
\end{center}
\end{figure}

In this Section we describe and characterize the dynamical behavior
of active rings once the steady state has been reached.
When $Pe \lesssim 1$, there are no effects induced by the
applied tangential force and rings behave as in the passive case with
diffusive translational motion of the center of mass (see the following
discussion).
By increasing activity, rings are set in a slow
rotational motion \textcolor{black}{due to} the applied force though this rotation
is not continuous in time. 
\textcolor{black}{In order to illustrate and quantify the described behavior,
it is useful to consider}
the ring diameter, defined as ${\mathbf R}_d={\bf r}_{N/2+1}-{\bf r}_{1}$.
\textcolor{black}{The time dependence of the $x$-component
$R_{dx}$ is reported in Fig.~\ref{fig:rdx} in the case of
a flexible self-avoiding ring at different values of activity.}
Once $Pe \sim o(10^2)$, a steady rotation
of active rings can be observed.
\textcolor{black}{During steady rotation, the vector ${\mathbf R}_d$}
rotates 
continuously so that its components oscillate periodically in time.
This behavior can be used to infer 
the characteristic rotation frequency \textcolor{black}{$\omega$}.
\textcolor{black}{This is determined by a spectral analysis
(see the inset of Fig.~\ref{fig:rdx} (d)) of the time
series $R_{dx}(t)$.}
The computed periods of rotation, \textcolor{black}{$T=2 \pi/\omega$},
are shown in Fig.~\ref{fig:period}
for different persistence lengths
 \textcolor{black}{and rings of lengths $L=50 l, 100 l$,
 \textcolor{black}{$200 l$}}.
 It is evident that
the period $T$ follows a power-law
decay with dependence \textcolor{black}{$(Pe/L^3)^{-1}$},
irrespective
of the bending rigidity
\textcolor{black}{and ring average size at high activity.
Our results confirm what analytically
predicted
for three-dimensional phantom active rings
that undergo active tank-treading motion
with frequency $\omega = (Pe/L^3) (2 \pi l k_B T/\xi)$
\textcolor{black}{$=f^a/(R_c \xi)$ which is
proportional to the tangential velocity $f^a/\xi$ and independent
of the effective ring size.}} \cite{phil:22}.
Moreover,
here we find evidence that the period is not depending on excluded volume
interactions in two dimensions.
{\black{In the case of the phantom flexible chain, a compact conformation is
observed at $Pe \simeq 10^2$ and
}}
thermal noise
deeply influences ring rotation {\black{so that}}
the observed
spectrum of frequencies is quite broad.
Phantom active rings require larger values of activity or of stiffness
with respect to  self-excluding active rings
in order to establish a uniform rotational motion.

\begin{figure}[ht]
\begin{center}
\includegraphics*[width=.99\textwidth,angle=0]{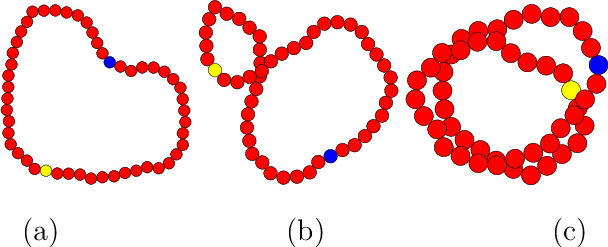}
\caption{Typical conformations of a ring
with $L_p/L=0.4$ at $Pe=2.5 \times 10^4$ for the PAR model
corresponding to the positions $R_g/R_c \simeq 0.95$ (a), $0.80$ (b),
$0.52$ (c) of the three peaks of the corresponding
probability distribution of  Fig.~\ref{fig:pdfrg_pe} (b).
The beads $1$ and $N/2+1$ are colored blue and yellow, respectively.
\label{fig:conf}
}
\end{center}
\end{figure}

\begin{figure}[ht]
\begin{center}
\includegraphics*[width=.9\textwidth,angle=0]{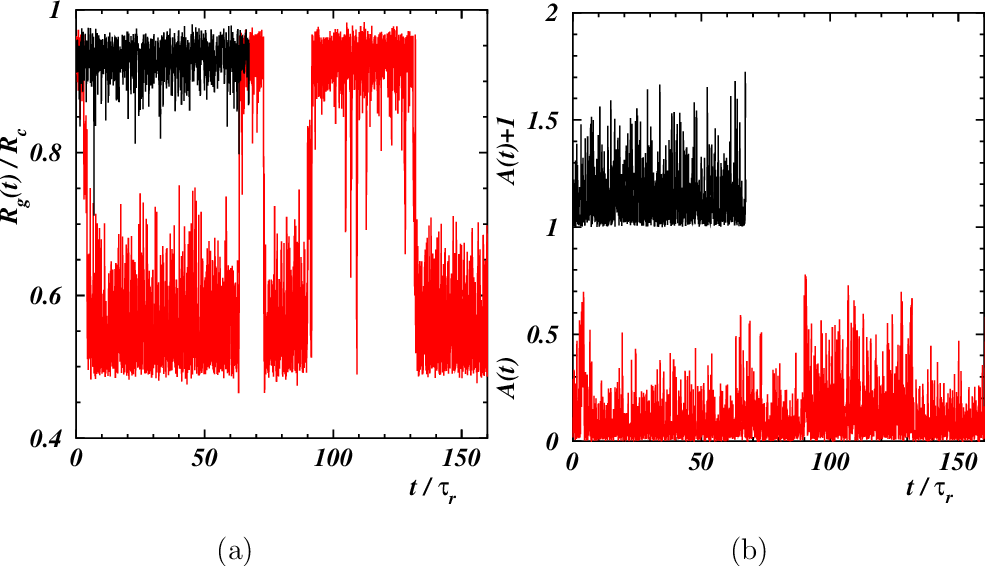}
\caption{Radius of gyration $R_g$ (a) and asphericity $A$ (b)
as functions of time $t/\tau_r$,
\textcolor{black}{where $\tau_r$ is the polymer relaxation time},
with $L_p/L=0.4$ at $Pe=2.5 \times 10^4$ for SAR (black line) and PAR (red line)
models. 
$R_c=L/(2 \pi)$ is the radius of
the corresponding rigid ring. The asphericity of the SAR model in panel (b)
has been shifted by $1$ to avoid overlap with the red curve.
\label{fig:rgirvst}
}
\end{center}
\end{figure}

Sizes and shapes of active rings in the steady
state show a poor dependence on the applied force as far as $Pe \lesssim 10^4$,
as already discussed in the previous Section. However,
when entering the regime of experimentally relevant P\'eclet numbers,
rings undergo large morphological deviations with respect to equilibrium.
Phantom active rings, despite the initial circular configuration,
can be driven, going through intermediate structures
(see panel (b) of Fig.~\ref{fig:conf}), into more compact
configurations (see panel (c) of Fig.~\ref{fig:conf}).
\textcolor{black}{Simulations for the PAR model have been conducted
at $Pe=2.5 \times 10^4$ for different values of the persistence length.
It appears that when $0.3 \lesssim L_p/L \lesssim 0.45$, rings
spontaneously assume the double ring conformation with
$R_g/R_c \simeq 0.52$ (corresponding to the typical value of $R_g$ 
for the conformation of Fig.~\ref{fig:conf} (c)).
This latter structure can spontaneously disentangle with a lifetime which
is longer
at $L_p/L \simeq 0.4$.
This behavior can be observed in
the time dependence of the gyration radius and of the asphericity
in Fig.~\ref{fig:rgirvst} for the PAR model
with $L_p/L=0.4$ at $Pe=2.5 \times 10^4$ on a very long time run of
duration $150 \tau_r \simeq 7 \times 10^4 T$.
Starting from the initial circular shape,
phantom rings can self-cross assuming conformations similar to the one
of Fig.~\ref{fig:conf} (b) with an elongated shape resembling the number eight.
This is possible only in a narrow range centered at $L_p/L \simeq 0.4$
since the ``eight configuration'' is compatible with this value of the
persistence length.
Due to thermal fluctuations, it can happen that one of the two sub-rings
moves towards the other one trespassing the mutual crossing point
to give the double ring conformation.
Despite this costs a strong local bending, the double ring is always
observed at $L_p/L=0.4$ in all the considered
runs at very high P\'eclet number.}
In the case of active rings comprising excluded volume interactions,
activity is responsible of inducing temporary elongated configurations
as illustrated in
Fig.~\ref{fig:rgirvst} by the
peaks of asphericity corresponding to the reduction
of the radius of gyration (see also Movie 2 in the supplementary material).

\begin{figure}[ht]
\begin{center}
\includegraphics*[width=.9\textwidth,angle=0]{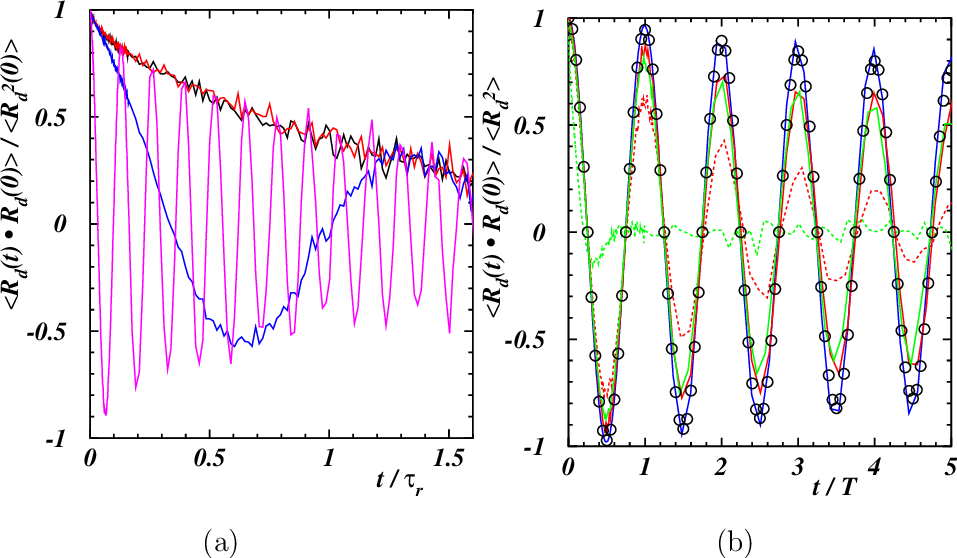}
\caption{(a) Autocorrelation function of the ring diameter ${\bf R}_d(t)$
as a function of time $t/\tau_r$, where $\tau_r$ is the polymer
relaxation time, for a self-avoiding active ring with $L_p/L=0.04$
and $Pe=0$ (black line), $10$ (red line), $10^2$ (blue line),
$10^3$ (purple line).
(b) Autocorrelation function of the ring diameter ${\bf R}_d(t)$
as a function of time $t/T$, where $T$ is the computed rotational period,
for $Pe=10^3$ and $L_p/L=0$ (green line), $0.1$ (red line), $40$ (blue line).
Full and dashed
lines correspond to the SAR and PAR models, respectively.
{\black{The black empty circles correspond
to the theoretical prediction (\ref{autocorr})
\cite{phil:22} when $L_p/L=40$.}}
\label{fig:re_bis}
}
\end{center}
\end{figure}

In order to further characterize the rotational behavior,
it is useful to consider
the normalized time-correlation function of the ring diameter
$\langle {\bf R}_d(t) \cdot {\bf R}_d(0) \rangle /
\langle {\bf R}_{d}^2(0) \rangle$.
In the left panel of Fig.~\ref{fig:re_bis} the normalized
autocorrelation function is plotted for a flexible self-avoiding 
ring for different values of activity.
In the passive case, the function exhibits an exponential decay,
$\exp(-t/\tau_r)$, which
is used  to determine the polymer relaxation time $\tau_r$.
When $Pe=10$ no relevant difference can be appreciated with respect to
equilibrium on time scales comparable to the relaxation time.
The increase of activity is responsible for producing an oscillatory
behavior which is modulated in time by the same decay of the passive ring.
The damped oscillatory pattern with a shorter period
is maintained when the P\'eclet is further increased.  
The comparison in the behavior of the autocorrelation function of the ring
diameter between the PAR and SAR models is reported in the panel (b)
of Fig.~\ref{fig:re_bis}
for different bending rigidities with $Pe=10^3$.
In the case of flexible phantom active rings, the correlation function
shows an exponential decay since the \textcolor{black}{observed compact}
structure, due to the lack
of any bending rigidity, requires larger values of activity to observe
oscillations. 
On the contrary, self-avoiding active rings present
the damped oscillatory behavior thanks to excluded volume effects that
preserve the circular topology avoiding any collapse of the chain
while rotating. 
Oscillations are clearly observable in the correlation functions of
semiflexible, both phantom and self-excluding, active rings.
\textcolor{black}{The amplitudes are larger in the latter case
due to the longer relaxation times
and increase with bending rigidity to
become indistinguishable between the two models in the limit of stiff rings.
As far as
oscillations are well defined, the numerical data of
the autocorrelation function are very well approximated
(see Fig.~\ref{fig:re_bis} (b)) by
the theoretical prediction \cite{phil:22}
\begin{equation}
\frac{\langle {\bf R}_d(t) \cdot {\bf R}_d(0) \rangle}
{\langle {\bf R}_{d}^2(0) \rangle} \approx \cos(2 \pi t / T) 
\exp(-t/\tau_r) ,
\label{autocorr}
\end{equation}
where the values of $T$ and $\tau_r$, computed in the present simulations,
are used.}

\begin{figure}[ht]
\begin{center}
\includegraphics*[width=.9\textwidth,angle=0]{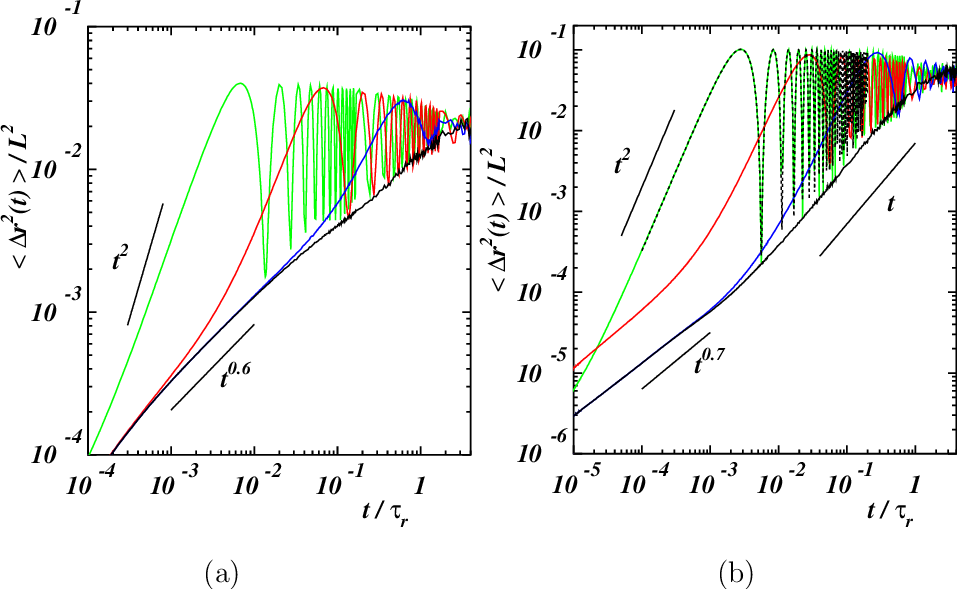}
\caption{{\black{Beads}} mean-square displacement {\black{relative to the}}
center-of-mass {\black{mean-square displacement}},
$\lla \Delta r^2(t) \rra$, of self-avoiding (a) flexible, $L_p/L=0$, and
(b) stiff, $L_p/L=40$,
active rings as
function of time $t/\tau_r$, where $\tau_r$ is the polymer
relaxation time,
for P\'eclet numbers $Pe=0$ (black line),
$10^2$ (blue line), $10^3$ (red line), $10^4$ (green line).
Short black lines indicate power laws with the annotated time dependence.
{\black{In panel (b) the black dashed line corresponds
to the theoretical prediction (\ref{msd})
\cite{phil:22} when $Pe=10^4$.}}
\label{fig:cmmsd}
}
\end{center}
\end{figure}

Finally, \textcolor{black}{the beads mean-square displacement
(MSD) $\lla ({\bf r}_i(t)-{\bf r}_i(0))^2 \rra$ is computed which allows
the characterization of the translational motion of ring. Due to the
ring topology, the beads MSD is independent of the point location
and receives a contribution from the center-of-mass motion,
$\lla \Delta {\bf r}_{cm}^2(t) \rra$, and another one from the internal
dynamics, $\lla \Delta {\bf r}^2(t) \rra$, so that one can write
$\lla ({\bf r}_i(t)-{\bf r}_i(0))^2 \rra=\lla \Delta {\bf r}_{cm}^2(t) \rra+
\lla \Delta {\bf r}^2(t) \rra$. 
Since} the sum of
all internal and active forces over the whole ring vanish,
the center-of-mass motion is purely diffusive
depending only on thermal fluctuations
and not on activity.
In this way the quantity $\lla \Delta {\bf r}^2(t) \rra$,
\textcolor{black}{which is related the beads MSD relative to the center-of-mass
MSD},
provides
information on the ring internal \textcolor{black}{dynamics}.
The MSD $\lla \Delta {\bf r}^2(t) \rra$ for self-avoiding flexible ($L_p/L=0$)
and stiff active rings ($L_p/L=40$)
with different activities are reported in Fig.~\ref{fig:cmmsd}.
In the case without any stiffness \textcolor{black}{(panel (a))}
the sub-diffusive exponent
$0.6$ is found in the time range $t \ll \tau_r$ when thermal effects
prevail on active contributions,
as predicted by the Rouse model of two-dimensional flexible polymers
with excluded volume interactions \cite{niko:16}.
For large P\'eclet numbers, $Pe \gtrsim 10^4$, an active ballistic time regime
is observed with $\lla \Delta {\bf r}^2(t) \rra \sim t^2$.
For longer times, oscillations, due to the active tank-treading, appear
in the MSD which then goes to a plateau when $t \gtrsim \tau_r$.
\textcolor{black}{This behavior, due to the mutual repulsion among beads,
is different from what is found
when considering flexible phantom rings.
In this case the sub-diffusive behavior
$t^{1/2}$ holds when $t \ll \tau_r$.
The MSD shows the
activity-enhanced linear time
regime at high values of activity ($Pe \simeq 10^4$) followed by oscillations
at longer times, as predicted in three dimensions \cite{phil:22}.}
The MSD of stiff polymers \textcolor{black}{(panel (b))}
exhibits an initial time dependence
\textcolor{black}{$t^{0.7}$. The exponent $0.7$ slightly underestimates
the predicted value $3/4$ \cite{farg:93} due to the finite
ring length \cite{niko:16}.
A linear time dependence \cite{wink:07.1} is then observed 
at late times when $Pe \lesssim 1$.}
Strong activity induces the active ballistic time regime
followed in time by oscillations.
\textcolor{black}{In this case we find that the numerical values of
$\lla \Delta {\bf r}^2(t) \rra$ are very well described
(see Fig.~\ref{fig:cmmsd} (b))
by the theoretical
prediction \cite{phil:22}
\begin{equation}
\lla \Delta {\bf r}^2(t) \rra / L^2 \approx
\Big [1-\cos(2 \pi t / T) e^{-t/\tau_r} \Big ] / (2 \pi^2) ,
\label{msd}
\end{equation}
where the computed values of
$T$ and $\tau_r$ are used.}

\section{Discussion and conclusions} \label{sec:conclusions}

The conformations and dynamics
of tangentially-driven active ring polymers
have been numerically studied.
The discrete closed chain has been confined in two dimensions and
coupled to a Brownian
heat bath performed by the stochastic implementation of the multi-particle
collision dynamics. 
Both phantom and self-avoiding rings
have been considered for different bending rigidities ranging from the flexible
to the stiff limit.

Excluded volume interactions affect the conformations of flexible active rings:
Polymers with self-excluding beads swell up to $15\%$ while phantom chains
shrink down to $25\%$ at high values of the P\'eclet number, in both cases
attaining more circular shapes. No appreciable difference is observed between
the two models of active rings in the semiflexible limit with the exception
of phantom rings with persistence length slightly less than
half of the total
chain length. In this latter case activity can induce more compact conformations
since the initial circular topology is not conserved.
\textcolor{black}{The observed double ring conformation
would be permitted in the presence of excluded-volume
interactions in three spatial dimensions, or
quasi two-dimensional confinement, so that activity might trigger
the transition to this conformation in real systems,
eventually detectable in experimental setups.}
The mean-square displacement \textcolor{black}{relative to the center-of-mass MSD}
allows us to capture the internal dynamics.
At intermediate time scales, flexible phantom active rings shows
an activity-enhanced diffusive regime
at large activity values. This is different from what observed
in the case of self-avoiding active chains for which internal motion is
ballistic, independently on stiffness.
\textcolor{black}{At high values of activity, rings exhibit active tank-treading
motion}
whose period scales as the inverse of the applied tangential force, irrespective
of both persistence length and beads self-exclusion.

Excluded volume interactions have been shown to play a major role
in capturing the phenomenology of two-dimensional flexible
active rings thus it appears
very interesting to extend the present study to melts where both
inter- and intra-bead mutual repulsions will be relevant. Moreover,
the action of an
external shear flow would possibly enrich the described picture
impacting on the observed conformations and dynamics,
as observed for active Brownian linear polymers \cite{mart:18,pand:23}. This
would require to separate out effects of internal and external stresses,
as far as timescales are not separable, in order to obtain a complete
description of the system.

\acknowledgments

Funding from MIUR Project No. PRIN 2020/PFCXPE is acknowledged.
This work was performed under the auspices of GNFM-INdAM.

\appendix*

\section{Movies Description}
In this Section we provide a brief description of the movies accompanying
the paper.
\begin{itemize}
\item{\textbf{Movie 1: Flexible self-avoiding active ring}\\
The movie
illustrates the motion
of a flexible self-avoiding active ring with $L_p/L=0$
for $Pe=2.5 \times 10^4$ in the center-of-mass reference frame.
Frames are taken at
time intervals $\Delta t / T \simeq 0.11$ where $T$ is the computed
rotational period. To illustrate
the clockwise rotation, the beads $1$ and $N/2+1$ are colored blue and yellow,
respectively.}
\item{\textbf{Movie 2: Semiflexible self-avoiding active ring}\\
The movie illustrates the motion
of a semiflexible self-avoiding active ring with $L_p/L=0.4$ for
$Pe=2.5 \times 10^4$ in the center-of-mass reference frame.
Frames are taken at
time intervals $\Delta t / T \simeq 0.11$ where $T$ is the computed
rotational period. To illustrate
the clockwise rotation, the beads $1$ and $N/2+1$ are colored blue and
yellow, respectively.}

\end{itemize}


%

\newpage
\clearpage

\newpage
\clearpage

\newpage
\clearpage

\newpage
\clearpage

\newpage
\clearpage

\newpage
\clearpage

\newpage
\clearpage

\newpage
\clearpage

\newpage
\clearpage

\newpage
\clearpage

\newpage
\clearpage

\end{document}